\documentclass[%
 reprint,
 amsmath,amssymb,
 aps,
]{revtex4-2}

\usepackage{graphicx}% Include figure files
\usepackage{dcolumn}% Align table columns on decimal point
\usepackage{bm}% bold math
\usepackage{comment}
\usepackage{color}
\begin{document}

\title{Nanofiber-induced losses inside an optical cavity}

\author{Bernd Welker}
\author{Thorsten Österle}
\author{Sebastian Slama}
\email{sebastian.slama@uni-tuebingen.de}
\affiliation{Center for Quantum Science and Physikalisches Institut, Eberhard-Karls Universität Tübingen, Auf der Morgenstelle 14, 72076 Tübingen, Germany }

\author{Thomas Hoinkes}
\author{Arno Rauschenbeutel}
\affiliation{Department of Physics, Humboldt-Universität zu Berlin, 10099 Berlin, Germany}

\date{\today}

\begin{abstract}
Optical high-finesse cavities are a well-known mean to enhance light-matter interactions. Despite large progress in the realization of strongly coupled light-matter systems, the controlled positioning of single solid emitters in cavity modes remains a challenge. We pursue the idea to use nanofibers with sub-wavelength diameter as a substrate for such emitters. This paper addresses the question how strongly optical nanofibers influence the cavity modes. We analyze the influence of the fiber position for various fiber diameters on the finesse of the cavity and on the shape of the modes.
\end{abstract}

\maketitle

%\tableofcontents

\section{\label{sec:intro} Introduction}
Strong coupling between light and matter is a hallmark of light-matter interactions and has been demonstrated in various systems using optical cavities\cite{Dovzhenko18}. It is reached when the coupling constant $g_0\propto 1/\sqrt{V_\mathrm{m}}$ with cavity mode volume $V_\mathrm{m}$ is larger than the excited state linewidth $\gamma$ of the emitter and the cavity linewidth $\kappa$ (the light field decay rate). Two complementary approaches exist in the setup of these cavities in order to enhance the coupling. One approach minimizes $V_\mathrm{m}$ by building micro- and nanocavities, for instance with photonic crystal stuctures \cite{Yoshie04, Englund10}, micropillars \cite{Reithmaier04}, microdisks \cite{Peter05,Barclaya09}, arrays of nanovoids \cite{Sugawara06}, and plasmonic nanostructures \cite{Zengin15, Chikkaraddy16}. The reduction of mode volume is connected with a reduction of the cavity round-trip length $L$, which in turn increases the free spectral range $\nu_{fsr}=\frac{c}{L}$ of the cavity and, correspondingly, the cavity full-width at half maximum $\nu_\mathrm{fwhm}=\frac{\kappa}{\pi}=\frac{\nu_{fsr}}{F}$ for given cavity finesse $F$. The finesse in nanocavities is limited by the unavoidable optical loss due to the optical properties of the materials (in particular of metals) and due to imperfections in the fabrication process. Moreover, emitters cannot be easily inserted or exchanged in nanoscale cavities. The second approach focuses on maximizing the finesse of the cavity by using superpolished mirrors made of dielectric layers with low optical loss. The best cavities can reach a finesse on the order of $F\sim 0.5\cdot 10^6$ \cite{Ye99}. They are typically coupled with ultracold atoms that are trapped in optical or magnetic traps in vacuum and can be positioned relative to the cavity mode with high precision. This is more complicated for solid emitters which are typically embedded in a dielectric substrate in the cavity \cite{Chizhik09} or are placed directly on one of the cavity mirrors \cite{Mader15}. The existence of a substrate limits the finesse due to additional optical losses by reflection at the substrate boundaries. Following the trapping of cold atoms, impressive progress has been made on the levitation and motional cooling of nanoparticles using Paul traps \cite{Dania21} and optical tweezers \cite{Delic20}. However, this approach is not applicable for arbitrary particles and requires extensive use of optics and servo loops.\\
\begin{figure}[t]
	\includegraphics{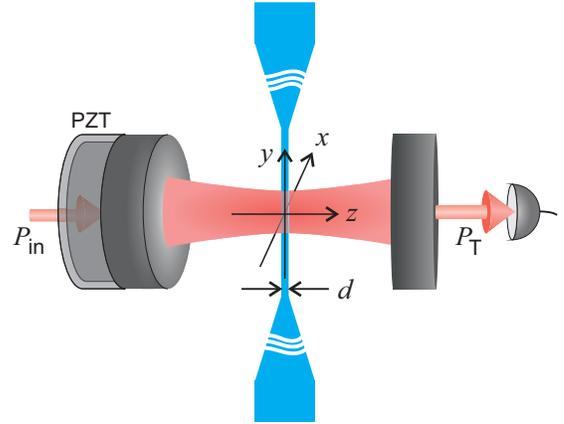}
	\caption{\label{fig1} (not to scale) A nanofiber with diameter $d$ is moved in the $x$-direction transversally through the mode of a cavity with free spectral range of $\nu_{fsr}=2.34~\mathrm{GHz}$ and finesse $F_0=1240$ (value without fiber). One of the mirrors is attached to a piezoelectric transducer (PZT) for scanning the cavity length. For fixed $x$-position of the nanofiber the cavity linewidth $\nu_\mathrm{fwhm}$ is determined from the cavity transmission.}
\end{figure}
This paper investigates a new substrate-based approach where the former plane substrate is replaced by an optical nanofiber. The latter are fabricated from standard step-index optical fibers in a heat-and-pull process, yielding a tapered optical fiber (TOF) with a subwavelength-diameter waist \cite{Warken08}. Optical emitters can be attached to these nanofibers, as demonstrated for nanocrystals \cite{Fujiwara11,Yalla12, Skoff18}, nitrogen-vacancy centers in diamond \cite{Yonezu17}, and gold nanospheres \cite{Ding20}. Since the TOF allows one to efficiently couple light into and out of the nanofiber waist, it can serve not only as a substrate but also as a fiber-optical interface, which allows one to efficiently excite the emitter and to collect its fluorescence. Thanks to the subwavelength-diameter, optical losses in the cavity due to the nanofiber are expected to be small. We analyze these losses by measuring the finesse of the cavity while nanofibers with various diameters are moved transversally through the mode. A similar setup has recently been used for measuring the optomechanical interaction between a nanowire and a cavity mode \cite{Fogliano21}.

\section{\label{sec:experiment} Experimental setup}
\begin{figure}[t]
\includegraphics{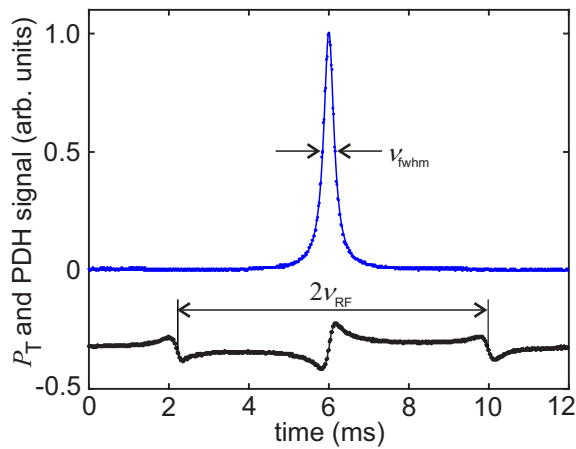}
\caption{\label{fig2} The cavity length is scanned across the resonance of a $\mathrm{TEM}_{00}$-mode. The cavity full width at half maximum $\nu_\mathrm{fwhm}$ is determined from the measured cavity transmission $P_T$ (blue data points). The frequency reference is given by the Pound-Drever-Hall signal (black data points)}
\end{figure}
The experimental setup is schematically drawn in Fig.~\ref{fig1}. An optical cavity is formed by two spherical dielectric mirrors with radius of curvature $R_c=6~\mathrm{cm}$ and intensity transmission $T\approx 0,25\%$, each. One of the mirrors is attached to a piezoelectric transducer (PZT). The free spectral range $\nu_{fsr}=2.34~\mathrm{GHz}$ of the cavity is calculated from the distance of the two mirrors of $L/2=6.4~\mathrm{cm}$. The beam waist ($1/e$-intensity radius) in the center of the cavity is $w_0=86~\mu m$. A laser field with power $P_\mathrm{in}= 3~\mathrm{mW}$ is coupled into the cavity through one of the mirrors. The laser field is generated by a free-running external cavity diode laser with wavelength $\lambda=780~\mathrm{nm}$ and $100~\mathrm{kHz}$ laser linewidth. Sidebands are generated by modulating the laser current with $\nu_\mathrm{RF}=22.1184~\mathrm{MHz}$ frequency in order to provide a frequency reference for the cavity spectrum. The spectrum is recorded by scanning the length of the cavity with the PZT, and monitoring the cavity transmission $P_T$ on a photodiode. A typical scan over the $\mathrm{TEM}_{00}$ mode is shown in Fig.~\ref{fig2}. The cavity linewidth is determined from a fit of the measured transmission. The distance between the two sidebands defines the frequency scale. However, they cannot be observed in the transmission signal at the same scale as the carrier. For that reason we simultaneously monitor a Pound-Drever-Hall signal in reflection from the cavity \cite{Drever83}, from which we determine the frequency scale.\\ 
TOFs with waist-diameters of $d=150~\mathrm{nm}$, $300~\mathrm{nm}$, $500~\mathrm{nm}$, and  $1000~\mathrm{nm}$ have been fabricated in a heat-and-pull process using a home-built computer-controlled fiber-pulling rig \cite{Warken07}. The TOF is then fixed on a mechanical mount, which we attach to a nanopositioning stage (Attocube ECS3030), so that it can be moved with high precision both in the $x$- and in the $z$-direction.   

\begin{figure}[b]
\includegraphics{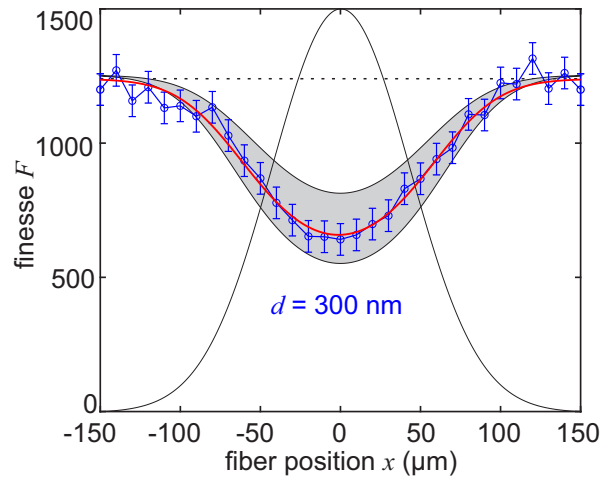}
\caption{\label{fig3} Experimentally determined finesse of the cavity (blue data) for different fiber positions $x$ of the $d=300~\mathrm{nm}$ thick fiber relative to the $\mathrm{TEM}_{00}$-mode (Gaussian lineshape). The dashed black line with a finesse of $F_0=1240$ corresponds to the empty cavity with mirror intensity transmission $T=0.253\%$. Error bars indicate the statistical variation of the determined finesse. Our model (red solid line) quantitatively agrees with the data. The grey shaded region indicates the accessible range by changing the (linear) light polarization.}
\end{figure}
\section{\label{sec:results} Results}
The finesse has been analyzed as a function of the fiber position $x$ transverse to the cavity mode. The result for the $300~\mathrm{nm}$ thick fiber is shown in Fig.~\ref{fig3}. It is reduced from a value of $F_0=1240$ without fiber to a minimum value of $F=650$ when the fiber is in the center of the mode. No finesse variation could be detected instead when the fiber was displaced in the $z$-direction along the cavity axis, where the mode function is given by the optical standing wave. In order to probe the field modulation of the standing wave the fiber must be orthogonal to the mode axis with an angle deviation of $\alpha\lesssim\frac{\lambda/2}{2 w_0}=0.13^\circ$. This was not the case in the present setup. The effective loss was given by an average over different positions along $z$.\\
The fiber-induced optical loss in the cavity is simulated as light scattering of the incident mode field from a circular cylinder (the fiber) following \cite{Hulst81}. The mode is simulated as two counter propagating plane waves with perpendicular incidence onto the fiber
\begin{equation}\label{eq:incident_wave}
u_\mathrm{inc}=u_0 e^{-ikz+i\omega t}+u_0 e^{ikz+i\omega t+i\phi},
\end{equation}
where the phase $\phi$ defines the position in the standing wave. Each incident wave generates a scattered wave of form 
\begin{equation}\label{eq:scattered_wave}
u(r,\theta)=\sqrt{\frac{2}{\pi k r}}u_0 e^{-ikr+i\omega t-i3\pi/4}T_{s,p}(\theta),
\end{equation}
 valid in the far field ($kr\gg d$), with distance $r$ from the fiber and angle deviation $\theta$ from the incident $k$-vector, both given in the $xz$ plane. The scattered field depends on the polarization (s or p) of the incident light field relative to the fiber and is invariant in the $y$-direction. The angle dependence is contained in the amplitude functions
\begin{eqnarray}\label{eq:amplitude_functions}
T_s(\theta)=b_0+2\sum_{n=1}^\infty b_n \cos(n \theta),\\
T_p(\theta)=a_0+2\sum_{n=1}^\infty a_n \cos(n \theta),
\end{eqnarray}
where the coefficients are given by Bessel functions $J_n(z)$ of first kind and second Hankel functions $H_n(z)$ and their derivatives $J'_n(z)$ and $H'_n(z)$ via
\begin{eqnarray}\label{eq:coefficients}
b_n=\frac{m J_n(y) J_n(x)-J_n(y) J'_n(x)}{m J'_n(y) H_n(x)-J_n(y) H'_n(x)},\\
a_n=\frac{J'_n(y) J_n(x)-m J_n(y) J'_n(x)}{J'_n(y) H_n(x)-m J_n(y) H'_n(x)}.
\end{eqnarray}
In these equations, $x=kd$ and $y=mkd$, and $m$ denotes the refractive index of the fiber. The scattered total field is given by the superposition of the amplitude functions in opposite directions
\begin{equation}\label{eq:total_amplitude_function}
T_{s,p}^\mathrm{tot}(\theta)=T_{s,p}(\theta)+e^{i\phi}T_{s,p}(\pi-\theta),
\end{equation}
corresponding to the scattered fields from the two counter propagating incident fields. The total power scattered by a fiber at position $x$ is calculated via integration of the scattered intensity around and along the fiber:
\begin{equation}\label{eq:total_scattered power}
P_\mathrm{scatt,s,p}=\int_0^{2\pi}rd\theta\int_{-\infty}^{+\infty}dy\left[\frac{2}{\pi kr}\left|T_{s,p}^\mathrm{tot}(\theta)\right|^2 I(x,y)\right],
\end{equation}
which is proportional to the incident intensity $I_0$ with Gaussian profile
\begin{equation}\label{eq:gaussian_intensity}
I(x,y)=I_0 e^{-2\frac{x^2+y^2}{w_0^2}},
\end{equation}
and thus to the light power
\begin{equation}\label{eq:power_cavity}
P_\mathrm{cav}=\frac{1}{2}\pi w_0^2I_0
\end{equation}
circulating in the cavity. The power loss per cavity round-trip caused by the fiber is given by
\begin{equation}\label{eq:loss}
L_\mathrm{s,p}=P_\mathrm{scatt,s,p}/P_\mathrm{cav}.
\end{equation}
This definition neglects scattering into resonant modes of the cavity not contributing to cavity loss which is a valid approximation for the beam divergence of the cavity mode in our experiment. The finesse of the cavity is connected with the fiber loss via
\begin{equation}\label{eq:finesse}
F=\frac{\pi\sqrt{r_m}}{1-r_m},
\end{equation}
with round-trip field reflectivity
\begin{equation}\label{eq:round_trip_refl}
r_m=\sqrt{1-L-T_1-T_2}.
\end{equation}
It depends periodically on the position $z$ of the fiber along the optical standing wave and on the light polarization. As the fiber in the experiment is slightly tilted with respect to the standing wave we take the average of the finesse over one period. The results for the two polarizations (s and p) define the upper and lower bound of the finesse enclosing the grey shaded area in Fig.\ref{fig3}. Experimentally, we couple circularly polarized light into the cavity. The sum of this incoming light field and the field reflected from the second cavity mirror leads to a linear polarization of the field in the cavity at an angle $\alpha$ with respect to the $y$-axis that depends on the position of the second cavity mirror. The effective loss 
\begin{equation}\label{eq:effective_loss}
L=\cos(\alpha)^2L_\mathrm{p}+\sin(\alpha)^2L_\mathrm{s}.
\end{equation}
takes this angle into account. The experimental data shown in Fig.~\ref{fig3} correspond to an angle of $\alpha=40^\circ$.\\
\begin{figure}[t]
\includegraphics{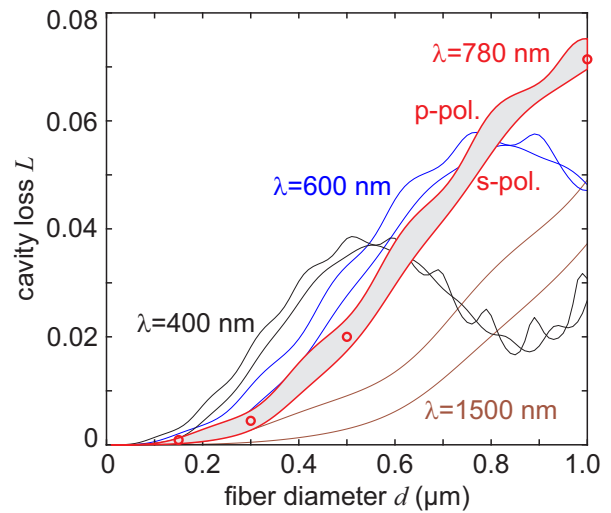}
\caption{\label{fig4} Optical loss in the cavity caused by a dielectric fiber of diameter $d$ in the center of the mode. Red circles are experimentally determined values, and solid lines are derived from the model for light polarization along (p) and perpendicular (s) to the fiber. For comparison, simulation results are presented for different optical wavelengths.}
\end{figure}
We now investigate how the cavity loss depends on the fiber diameter. Results are shown in Fig.\ref{fig4}. The red circles have been determined from the experimental data shown in Fig.\ref{fig3} and equivalent data for the other fiber diameters. They correspond to the maximum loss when the fiber is in the center of the mode. The experimental uncertainty is within the circle diameter. We see that the loss is strongly reduced for fiber diameters well below the optical wavelength of $780~\mathrm{nm}$. The data are consistent with the theoretical calculation using the model explained above. It is notable that the loss is slightly larger when the light field is polarized parallel to the fiber. Moreover, small oscillations are observed in the simulation that can be attributed to optical resonances in the fiber. These resonances are beyond the scope of this work. Fig.\ref{fig4} also illustrates how the losses depend on the optical wavelength. Shorter wavelengths lead to more pronounced oscillations. Larger wavelengths lead to smaller losses, at least for small fiber diameters.\\
\begin{figure}[t]
\includegraphics{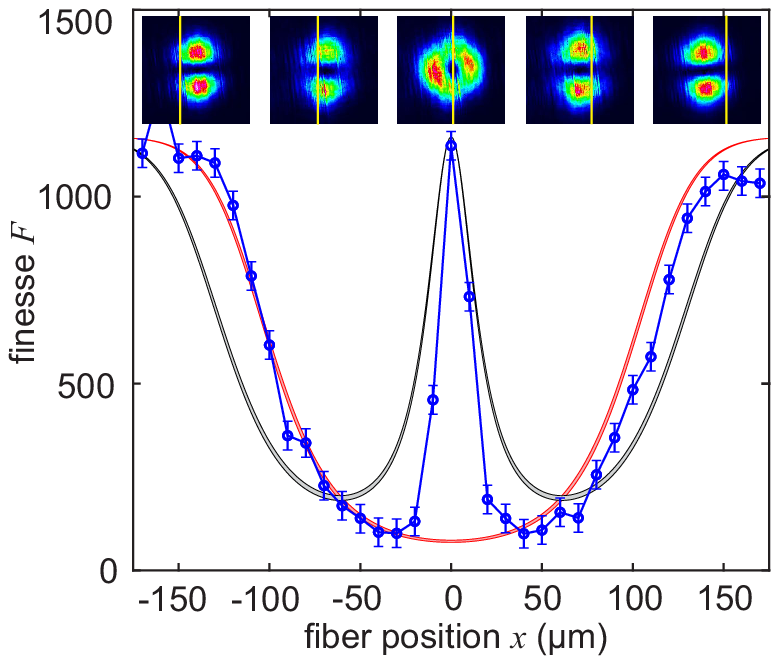}
\caption{\label{fig5} The finesse in a TEM$_{10}$ mode has a sharp maximum when the fiber is in the center of the mode. In this case, only the near-degenerate TEM$_{01}$ mode is excited whose node line is parallel to the fiber axis. The fiber diameter is $d=1~\mu m$. The mode shapes have been detected by imaging the light beam transmitted through the cavity with a camera. The approximate position of the fiber (not to scale) has been been drawn into the pictures. The red/black lines correspond to the loss rates  expected for a TEM$_{00}$/TEM$_{01}$ mode, respectively.}
\end{figure}

The situation is more complicated for higher transverse TEM modes of the cavity, where near-degenerate modes exist. In this case, the presence of the fiber influences which mode is excited. To illustrate this connection, we excite the TEM$_{10}$ mode which has a node line in the $x$-direction. Due to the cylindrical geometry of the cavity, the TEM$_{10}$ mode is near-degenerate with the TEM$_{01}$ mode which has a node line in the $y$-direction. As the fiber with its axis parallel to the $y$-direction is moved into the mode, the observed finesse first goes down, similar to the situation with the TEM$_{00}$ mode, see Fig.\ref{fig5}. However, in the very center of the mode the finesse increases again to a value comparable to the case without fiber. At the same time, the transverse intensity profile in the cavity turns around the cavity axis by $90^\circ$ where the fiber axis coincides with the node line of the TEM$_{01}$ mode. This observation can be understood as caused by a mode-dependant interaction: without nanofiber, both modes are excited depending on their geometric overlap with the incoming laser beam. In the situation of Fig.~\ref{fig5} the incoming laser beam is adjusted to mainly excite the TEM$_{10}$ mode, however also the TEM$_{01}$ mode is slightly excited (not visible in the image). As the nanofiber is introduced into the cavity the modes suffer both losses and a dispersive phase shift depending on the overlap of the respective mode intensity with the nanofiber. In the very center of the cavity where the overlap with the TEM$_{10}$ / TEM$_{01}$ mode is maximum / minimum, respectively, the different dispersive phase shifts lift the degeneracy of the two modes. Thus, the TEM$_{01}$ mode can be separately excited without loss for a certain laser frequency, whereas the TEM$_{10}$ mode which suffers loss is shifted to a different frequency. The data shown in Fig.\ref{fig5} were taken with the $d=1~\mu m$ diameter fiber where this effect was strongest. We observed that the mode rotation gets weaker for decreasing fiber diameter, consistent with a smaller dispersive shift of the TEM$_{10}$ mode. We will investigate the phase shift quantitatively in future work. Fig.~\ref{fig5} compares the experimental data with the results from our model. The red lines correspond to a loss rate as expected for the TEM$_{00}$ mode with no node line in the center, and the black lines correspond to a loss rate as expected for the TEM$_{01}$ mode with intensity profile
\begin{equation}\label{eq:TEM_01}
I(x,y)=I_0 e^{-2\frac{x^2+y^2}{w_0^2}}\left(\frac{\sqrt{2}x}{w_0}\right)^2.
\end{equation}

\section{\label{sec:conclusion} Conclusion}
This paper analyzes the optical losses in a cavity caused by light scattering from an optical nanofiber, and its dependence on the fiber diameter and polarization of the mode. For the smallest fiber diameter tested ($d=150~\mathrm{nm}$) the round-trip loss was on the order of $L=0.001$ such that the cavity finesse with fiber in the center of the mode reached a value on the order of $F=1000$. Thus, the fiber might be a very useful substrate to investigate the optical properties of single nanoparticles in high-finesse cavities. The fiber-induced loss could be reduced even further by positioning the fiber in a node of the standing wave in the cavity, if the fiber axis is aligned perpendicular to the cavity axis. An interesting feature was also observed for the degenerate TEM$_{10}$ and TEM$_{01}$ modes where the position of the fiber determined which mode was excited. An even richer behaviour can be expected for confocal or concentric cavities that facilitate many degenerate modes. Another interesting perspective is the possibility to couple the cavity modes to light propagating along the fiber as evanescent wave. A nanoparticle attached to the fiber could serve as a switchable interface between the fiber and cavity mode with applications in sensing and single photon generation. Another perspective of the setup is the investigation of quantum Mie scattering \cite{Maurer21}.  

%\begin{acknowledgments}
%We wish to acknowledge no-one and nothing.
%\end{acknowledgments}

%\begin{thebibliography}{10}
 
%\end{thebibliography}

%\begin{thebibliography}{10}
\bibliography{Slama}% Produces the bibliography via BibTeX.

\providecommand{\noopsort}[1]{}\providecommand{\singleletter}[1]{#1}%
\begin{thebibliography}{25}%
\makeatletter
\providecommand \@ifxundefined [1]{%
 \@ifx{#1\undefined}
}%
\providecommand \@ifnum [1]{%
 \ifnum #1\expandafter \@firstoftwo
 \else \expandafter \@secondoftwo
 \fi
}%
\providecommand \@ifx [1]{%
 \ifx #1\expandafter \@firstoftwo
 \else \expandafter \@secondoftwo
 \fi
}%
\providecommand \natexlab [1]{#1}%
\providecommand \enquote  [1]{``#1''}%
\providecommand \bibnamefont  [1]{#1}%
\providecommand \bibfnamefont [1]{#1}%
\providecommand \citenamefont [1]{#1}%
\providecommand \href@noop [0]{\@secondoftwo}%
\providecommand \href [0]{\begingroup \@sanitize@url \@href}%
\providecommand \@href[1]{\@@startlink{#1}\@@href}%
\providecommand \@@href[1]{\endgroup#1\@@endlink}%
\providecommand \@sanitize@url [0]{\catcode `\\12\catcode `\$12\catcode
  `\&12\catcode `\#12\catcode `\^12\catcode `\_12\catcode `\%12\relax}%
\providecommand \@@startlink[1]{}%
\providecommand \@@endlink[0]{}%
\providecommand \url  [0]{\begingroup\@sanitize@url \@url }%
\providecommand \@url [1]{\endgroup\@href {#1}{\urlprefix }}%
\providecommand \urlprefix  [0]{URL }%
\providecommand \Eprint [0]{\href }%
\providecommand \doibase [0]{https://doi.org/}%
\providecommand \selectlanguage [0]{\@gobble}%
\providecommand \bibinfo  [0]{\@secondoftwo}%
\providecommand \bibfield  [0]{\@secondoftwo}%
\providecommand \translation [1]{[#1]}%
\providecommand \BibitemOpen [0]{}%
\providecommand \bibitemStop [0]{}%
\providecommand \bibitemNoStop [0]{.\EOS\space}%
\providecommand \EOS [0]{\spacefactor3000\relax}%
\providecommand \BibitemShut  [1]{\csname bibitem#1\endcsname}%
\let\auto@bib@innerbib\@empty
%</preamble>
\bibitem [{\citenamefont {Dovzhenko}\ \emph {et~al.}(2018)\citenamefont
  {Dovzhenko}, \citenamefont {Ryabchuk}, \citenamefont {Rakovich},\ and\
  \citenamefont {Nabiev}}]{Dovzhenko18}%
  \BibitemOpen
  \bibfield  {author} {\bibinfo {author} {\bibfnamefont {D.~S.}\ \bibnamefont
  {Dovzhenko}}, \bibinfo {author} {\bibfnamefont {S.~V.}\ \bibnamefont
  {Ryabchuk}}, \bibinfo {author} {\bibfnamefont {Y.~P.}\ \bibnamefont
  {Rakovich}},\ and\ \bibinfo {author} {\bibfnamefont {I.~R.}\ \bibnamefont
  {Nabiev}},\ }\bibfield  {title} {\bibinfo {title} {Light-metter interactions
  in the strong coupling regime: configurations, conditions, and
  applications},\ }\href@noop {} {\bibfield  {journal} {\bibinfo  {journal}
  {Nanoscale}\ }\textbf {\bibinfo {volume} {10}},\ \bibinfo {pages} {3589}
  (\bibinfo {year} {2018})},\ \bibinfo {note}
  {{DOI:10.1039/c7nr06917k}}\BibitemShut {NoStop}%
\bibitem [{\citenamefont {Yoshie}\ \emph {et~al.}(2004)\citenamefont {Yoshie},
  \citenamefont {Scherer}, \citenamefont {Hendrickson}, \citenamefont
  {Khitrova}, \citenamefont {Gibbs}, \citenamefont {Rupper}, \citenamefont
  {Ell}, \citenamefont {Shchekin},\ and\ \citenamefont {Deppe}}]{Yoshie04}%
  \BibitemOpen
  \bibfield  {author} {\bibinfo {author} {\bibfnamefont {T.}~\bibnamefont
  {Yoshie}}, \bibinfo {author} {\bibfnamefont {A.}~\bibnamefont {Scherer}},
  \bibinfo {author} {\bibfnamefont {J.}~\bibnamefont {Hendrickson}}, \bibinfo
  {author} {\bibfnamefont {G.}~\bibnamefont {Khitrova}}, \bibinfo {author}
  {\bibfnamefont {H.~M.}\ \bibnamefont {Gibbs}}, \bibinfo {author}
  {\bibfnamefont {G.}~\bibnamefont {Rupper}}, \bibinfo {author} {\bibfnamefont
  {C.}~\bibnamefont {Ell}}, \bibinfo {author} {\bibfnamefont {O.~B.}\
  \bibnamefont {Shchekin}},\ and\ \bibinfo {author} {\bibfnamefont {D.~G.}\
  \bibnamefont {Deppe}},\ }\bibfield  {title} {\bibinfo {title} {Vacuum rabi
  splitting with a single quantum dot in a photonic crystal nanocavity},\
  }\href@noop {} {\bibfield  {journal} {\bibinfo  {journal} {Nature}\ }\textbf
  {\bibinfo {volume} {432}},\ \bibinfo {pages} {200} (\bibinfo {year}
  {2004})},\ \bibinfo {note} {{DOI:10.1038/nature03119}}\BibitemShut {NoStop}%
\bibitem [{\citenamefont {Englund}\ \emph {et~al.}(2010)\citenamefont
  {Englund}, \citenamefont {Shields}, \citenamefont {Rivoire}, \citenamefont
  {Hatami}, \citenamefont {Vučković}, \citenamefont {Park},\ and\
  \citenamefont {Lukin}}]{Englund10}%
  \BibitemOpen
  \bibfield  {author} {\bibinfo {author} {\bibfnamefont {D.}~\bibnamefont
  {Englund}}, \bibinfo {author} {\bibfnamefont {B.}~\bibnamefont {Shields}},
  \bibinfo {author} {\bibfnamefont {K.}~\bibnamefont {Rivoire}}, \bibinfo
  {author} {\bibfnamefont {F.}~\bibnamefont {Hatami}}, \bibinfo {author}
  {\bibfnamefont {J.}~\bibnamefont {Vučković}}, \bibinfo {author}
  {\bibfnamefont {H.}~\bibnamefont {Park}},\ and\ \bibinfo {author}
  {\bibfnamefont {M.~D.}\ \bibnamefont {Lukin}},\ }\bibfield  {title} {\bibinfo
  {title} {Deterministic coupling of a single nitrogen vacancy center to a
  photonic crystal cavity},\ }\href@noop {} {\bibfield  {journal} {\bibinfo
  {journal} {Nano Lett.}\ }\textbf {\bibinfo {volume} {10}},\ \bibinfo {pages}
  {3922–3926} (\bibinfo {year} {2010})},\ \bibinfo {note}
  {{DOI:10.1021/nl101662v}}\BibitemShut {NoStop}%
\bibitem [{\citenamefont {Reithmaier}\ \emph {et~al.}(2004)\citenamefont
  {Reithmaier}, \citenamefont {Sek}, \citenamefont {Löffler}, \citenamefont
  {Hofmann}, \citenamefont {Kuhn}, \citenamefont {Reitzenstein}, \citenamefont
  {Keldysh}, \citenamefont {Kulakovskii}, \citenamefont {Reinecke},\ and\
  \citenamefont {Forchel}}]{Reithmaier04}%
  \BibitemOpen
  \bibfield  {author} {\bibinfo {author} {\bibfnamefont {J.~P.}\ \bibnamefont
  {Reithmaier}}, \bibinfo {author} {\bibfnamefont {G.}~\bibnamefont {Sek}},
  \bibinfo {author} {\bibfnamefont {A.}~\bibnamefont {Löffler}}, \bibinfo
  {author} {\bibfnamefont {C.}~\bibnamefont {Hofmann}}, \bibinfo {author}
  {\bibfnamefont {S.}~\bibnamefont {Kuhn}}, \bibinfo {author} {\bibfnamefont
  {S.}~\bibnamefont {Reitzenstein}}, \bibinfo {author} {\bibfnamefont {L.~V.}\
  \bibnamefont {Keldysh}}, \bibinfo {author} {\bibfnamefont {V.~D.}\
  \bibnamefont {Kulakovskii}}, \bibinfo {author} {\bibfnamefont {T.~L.}\
  \bibnamefont {Reinecke}},\ and\ \bibinfo {author} {\bibfnamefont
  {A.}~\bibnamefont {Forchel}},\ }\bibfield  {title} {\bibinfo {title} {Strong
  coupling in a single quantum dot–semiconductor microcavity system},\
  }\href@noop {} {\bibfield  {journal} {\bibinfo  {journal} {Nature}\ }\textbf
  {\bibinfo {volume} {432}},\ \bibinfo {pages} {197} (\bibinfo {year}
  {2004})},\ \bibinfo {note} {{DOI:10.1038/nature02969}}\BibitemShut {NoStop}%
\bibitem [{\citenamefont {Peter}\ \emph {et~al.}(2005)\citenamefont {Peter},
  \citenamefont {Senellart}, \citenamefont {Martrou}, \citenamefont
  {Lemaître}, \citenamefont {Hours}, \citenamefont {Gérard},\ and\
  \citenamefont {Bloch}}]{Peter05}%
  \BibitemOpen
  \bibfield  {author} {\bibinfo {author} {\bibfnamefont {E.}~\bibnamefont
  {Peter}}, \bibinfo {author} {\bibfnamefont {P.}~\bibnamefont {Senellart}},
  \bibinfo {author} {\bibfnamefont {D.}~\bibnamefont {Martrou}}, \bibinfo
  {author} {\bibfnamefont {A.}~\bibnamefont {Lemaître}}, \bibinfo {author}
  {\bibfnamefont {J.}~\bibnamefont {Hours}}, \bibinfo {author} {\bibfnamefont
  {J.~M.}\ \bibnamefont {Gérard}},\ and\ \bibinfo {author} {\bibfnamefont
  {J.}~\bibnamefont {Bloch}},\ }\bibfield  {title} {\bibinfo {title}
  {Exciton-photon strong-coupling regime for a single quantum dot embedded in a
  microcavity},\ }\href@noop {} {\bibfield  {journal} {\bibinfo  {journal}
  {Phys. Rev. Lett.}\ }\textbf {\bibinfo {volume} {95}},\ \bibinfo {pages}
  {067401} (\bibinfo {year} {2005})},\ \bibinfo {note}
  {{DOI:10.1103/PhysRevLett.95.067401}}\BibitemShut {NoStop}%
\bibitem [{\citenamefont {Barclaya}\ \emph {et~al.}(2009)\citenamefont
  {Barclaya}, \citenamefont {Fu}, \citenamefont {Santori},\ and\ \citenamefont
  {Beausoleil}}]{Barclaya09}%
  \BibitemOpen
  \bibfield  {author} {\bibinfo {author} {\bibfnamefont {P.~E.}\ \bibnamefont
  {Barclaya}}, \bibinfo {author} {\bibfnamefont {K.-M.~C.}\ \bibnamefont {Fu}},
  \bibinfo {author} {\bibfnamefont {C.}~\bibnamefont {Santori}},\ and\ \bibinfo
  {author} {\bibfnamefont {R.~G.}\ \bibnamefont {Beausoleil}},\ }\bibfield
  {title} {\bibinfo {title} {Chip-based microcavities coupled to
  nitrogen-vacancy centers in single crystal diamond},\ }\href@noop {}
  {\bibfield  {journal} {\bibinfo  {journal} {Appl. Phys. Lett.}\ }\textbf
  {\bibinfo {volume} {95}},\ \bibinfo {pages} {191115} (\bibinfo {year}
  {2009})},\ \bibinfo {note} {{DOI:10.1063/1.3262948}}\BibitemShut {NoStop}%
\bibitem [{\citenamefont {Sugawara}\ \emph {et~al.}(2006)\citenamefont
  {Sugawara}, \citenamefont {Kelf}, \citenamefont {Baumberg}, \citenamefont
  {Abdelsalam},\ and\ \citenamefont {Bartlett}}]{Sugawara06}%
  \BibitemOpen
  \bibfield  {author} {\bibinfo {author} {\bibfnamefont {Y.}~\bibnamefont
  {Sugawara}}, \bibinfo {author} {\bibfnamefont {T.~A.}\ \bibnamefont {Kelf}},
  \bibinfo {author} {\bibfnamefont {J.~J.}\ \bibnamefont {Baumberg}}, \bibinfo
  {author} {\bibfnamefont {M.~E.}\ \bibnamefont {Abdelsalam}},\ and\ \bibinfo
  {author} {\bibfnamefont {P.~N.}\ \bibnamefont {Bartlett}},\ }\bibfield
  {title} {\bibinfo {title} {Strong coupling between localized plasmons and
  organic excitons in metal nanovoids},\ }\href@noop {} {\bibfield  {journal}
  {\bibinfo  {journal} {Phys. Rev. Lett.}\ }\textbf {\bibinfo {volume} {97}},\
  \bibinfo {pages} {266808} (\bibinfo {year} {2006})},\ \bibinfo {note}
  {{DOI:10.1103/PhysRevLett.97.266808}}\BibitemShut {NoStop}%
\bibitem [{\citenamefont {Zengin}\ \emph {et~al.}(2015)\citenamefont {Zengin},
  \citenamefont {Wersäll}, \citenamefont {Nilsson}, \citenamefont
  {Antosiewicz}, \citenamefont {Käll},\ and\ \citenamefont
  {Shegai}}]{Zengin15}%
  \BibitemOpen
  \bibfield  {author} {\bibinfo {author} {\bibfnamefont {G.}~\bibnamefont
  {Zengin}}, \bibinfo {author} {\bibfnamefont {M.}~\bibnamefont {Wersäll}},
  \bibinfo {author} {\bibfnamefont {S.}~\bibnamefont {Nilsson}}, \bibinfo
  {author} {\bibfnamefont {T.~J.}\ \bibnamefont {Antosiewicz}}, \bibinfo
  {author} {\bibfnamefont {M.}~\bibnamefont {Käll}},\ and\ \bibinfo {author}
  {\bibfnamefont {T.}~\bibnamefont {Shegai}},\ }\bibfield  {title} {\bibinfo
  {title} {Realizing strong light-matter interactions between
  single-nanoparticle plasmons and molecular excitons at ambient conditions},\
  }\href@noop {} {\bibfield  {journal} {\bibinfo  {journal} {Phys. Rev. Lett.}\
  }\textbf {\bibinfo {volume} {114}},\ \bibinfo {pages} {157401} (\bibinfo
  {year} {2015})},\ \bibinfo {note}
  {{DOI:10.1103/PhysRevLett.114.157401}}\BibitemShut {NoStop}%
\bibitem [{\citenamefont {Chikkaraddy}\ \emph {et~al.}(2016)\citenamefont
  {Chikkaraddy}, \citenamefont {de~Nijs}, \citenamefont {Benz}, \citenamefont
  {Barrow}, \citenamefont {Scherman}, \citenamefont {Rosta}, \citenamefont
  {Demetriadou}, \citenamefont {Fox}, \citenamefont {Hess},\ and\ \citenamefont
  {Baumberg}}]{Chikkaraddy16}%
  \BibitemOpen
  \bibfield  {author} {\bibinfo {author} {\bibfnamefont {R.}~\bibnamefont
  {Chikkaraddy}}, \bibinfo {author} {\bibfnamefont {B.}~\bibnamefont
  {de~Nijs}}, \bibinfo {author} {\bibfnamefont {F.}~\bibnamefont {Benz}},
  \bibinfo {author} {\bibfnamefont {S.~J.}\ \bibnamefont {Barrow}}, \bibinfo
  {author} {\bibfnamefont {O.~A.}\ \bibnamefont {Scherman}}, \bibinfo {author}
  {\bibfnamefont {E.}~\bibnamefont {Rosta}}, \bibinfo {author} {\bibfnamefont
  {A.}~\bibnamefont {Demetriadou}}, \bibinfo {author} {\bibfnamefont
  {P.}~\bibnamefont {Fox}}, \bibinfo {author} {\bibfnamefont {O.}~\bibnamefont
  {Hess}},\ and\ \bibinfo {author} {\bibfnamefont {J.~J.}\ \bibnamefont
  {Baumberg}},\ }\bibfield  {title} {\bibinfo {title} {Single-molecule strong
  coupling at room temperature in plasmonic nanocavities},\ }\href@noop {}
  {\bibfield  {journal} {\bibinfo  {journal} {Nature}\ }\textbf {\bibinfo
  {volume} {535}},\ \bibinfo {pages} {127} (\bibinfo {year} {2016})},\ \bibinfo
  {note} {{DOI:10.1038/nature17974}}\BibitemShut {NoStop}%
\bibitem [{\citenamefont {Ye}\ \emph {et~al.}(1999)\citenamefont {Ye},
  \citenamefont {Vernooy},\ and\ \citenamefont {Kimble}}]{Ye99}%
  \BibitemOpen
  \bibfield  {author} {\bibinfo {author} {\bibfnamefont {J.}~\bibnamefont
  {Ye}}, \bibinfo {author} {\bibfnamefont {D.~W.}\ \bibnamefont {Vernooy}},\
  and\ \bibinfo {author} {\bibfnamefont {H.~J.}\ \bibnamefont {Kimble}},\
  }\bibfield  {title} {\bibinfo {title} {Trapping of single atoms in cavity
  qed},\ }\href@noop {} {\bibfield  {journal} {\bibinfo  {journal} {Phys. Rev.
  Lett.}\ }\textbf {\bibinfo {volume} {83}},\ \bibinfo {pages} {4987} (\bibinfo
  {year} {1999})},\ \bibinfo {note}
  {{DOI:10.1103/PhysRevLett.83.4987}}\BibitemShut {NoStop}%
\bibitem [{\citenamefont {Chizhik}\ \emph {et~al.}(2009)\citenamefont
  {Chizhik}, \citenamefont {Schleifenbaum}, \citenamefont {Gutbrod},
  \citenamefont {Chizhik}, \citenamefont {Khoptyar}, \citenamefont {Meixner},\
  and\ \citenamefont {Enderlein}}]{Chizhik09}%
  \BibitemOpen
  \bibfield  {author} {\bibinfo {author} {\bibfnamefont {A.}~\bibnamefont
  {Chizhik}}, \bibinfo {author} {\bibfnamefont {F.}~\bibnamefont
  {Schleifenbaum}}, \bibinfo {author} {\bibfnamefont {R.}~\bibnamefont
  {Gutbrod}}, \bibinfo {author} {\bibfnamefont {A.}~\bibnamefont {Chizhik}},
  \bibinfo {author} {\bibfnamefont {D.}~\bibnamefont {Khoptyar}}, \bibinfo
  {author} {\bibfnamefont {A.~J.}\ \bibnamefont {Meixner}},\ and\ \bibinfo
  {author} {\bibfnamefont {J.}~\bibnamefont {Enderlein}},\ }\bibfield  {title}
  {\bibinfo {title} {Tuning the fluorescence emission spectra of a single
  molecule with a variable optical subwavelength metal microcavity},\
  }\href@noop {} {\bibfield  {journal} {\bibinfo  {journal} {Phys. Rev. Lett.}\
  }\textbf {\bibinfo {volume} {102}},\ \bibinfo {pages} {073002} (\bibinfo
  {year} {2009})},\ \bibinfo {note}
  {{DOI:10.1103/PhysRevLett.102.073002}}\BibitemShut {NoStop}%
\bibitem [{\citenamefont {Mader}\ \emph {et~al.}(2015)\citenamefont {Mader},
  \citenamefont {Reichel}, \citenamefont {Hänsch},\ and\ \citenamefont
  {Hunger}}]{Mader15}%
  \BibitemOpen
  \bibfield  {author} {\bibinfo {author} {\bibfnamefont {M.}~\bibnamefont
  {Mader}}, \bibinfo {author} {\bibfnamefont {J.}~\bibnamefont {Reichel}},
  \bibinfo {author} {\bibfnamefont {T.~W.}\ \bibnamefont {Hänsch}},\ and\
  \bibinfo {author} {\bibfnamefont {D.}~\bibnamefont {Hunger}},\ }\bibfield
  {title} {\bibinfo {title} {A scanning cavity microscope},\ }\href@noop {}
  {\bibfield  {journal} {\bibinfo  {journal} {Nat Commun}\ }\textbf {\bibinfo
  {volume} {6}},\ \bibinfo {pages} {7249} (\bibinfo {year} {2015})},\ \bibinfo
  {note} {{DOI:10.1038/ncomms8249}}\BibitemShut {NoStop}%
\bibitem [{\citenamefont {Dania}\ \emph {et~al.}(2021)\citenamefont {Dania},
  \citenamefont {Bykov}, \citenamefont {Knoll}, \citenamefont {Mestres},\ and\
  \citenamefont {Northup}}]{Dania21}%
  \BibitemOpen
  \bibfield  {author} {\bibinfo {author} {\bibfnamefont {L.}~\bibnamefont
  {Dania}}, \bibinfo {author} {\bibfnamefont {D.~S.}\ \bibnamefont {Bykov}},
  \bibinfo {author} {\bibfnamefont {M.}~\bibnamefont {Knoll}}, \bibinfo
  {author} {\bibfnamefont {P.}~\bibnamefont {Mestres}},\ and\ \bibinfo {author}
  {\bibfnamefont {T.~E.}\ \bibnamefont {Northup}},\ }\bibfield  {title}
  {\bibinfo {title} {Optical and electrical feedback cooling of a silica
  nanoparticle levitated in a paul trap},\ }\href@noop {} {\bibfield  {journal}
  {\bibinfo  {journal} {Phys. Rev. Research}\ }\textbf {\bibinfo {volume}
  {3}},\ \bibinfo {pages} {013018} (\bibinfo {year} {2021})},\ \bibinfo {note}
  {{DOI:10.1103/PhysRevResearch.3.013018}}\BibitemShut {NoStop}%
\bibitem [{\citenamefont {Delic}\ \emph {et~al.}(2020)\citenamefont {Delic},
  \citenamefont {Reisenbauer}, \citenamefont {Dare}, \citenamefont {Grass},
  \citenamefont {Vuletić}, \citenamefont {Kiesel},\ and\ \citenamefont
  {Aspelmeyer}}]{Delic20}%
  \BibitemOpen
  \bibfield  {author} {\bibinfo {author} {\bibfnamefont {U.}~\bibnamefont
  {Delic}}, \bibinfo {author} {\bibfnamefont {M.}~\bibnamefont {Reisenbauer}},
  \bibinfo {author} {\bibfnamefont {K.}~\bibnamefont {Dare}}, \bibinfo {author}
  {\bibfnamefont {D.}~\bibnamefont {Grass}}, \bibinfo {author} {\bibfnamefont
  {V.}~\bibnamefont {Vuletić}}, \bibinfo {author} {\bibfnamefont
  {N.}~\bibnamefont {Kiesel}},\ and\ \bibinfo {author} {\bibfnamefont
  {M.}~\bibnamefont {Aspelmeyer}},\ }\bibfield  {title} {\bibinfo {title}
  {Cooling of a levitated nanoparticle to the motional quantum ground state},\
  }\href@noop {} {\bibfield  {journal} {\bibinfo  {journal} {Science}\ }\textbf
  {\bibinfo {volume} {367}},\ \bibinfo {pages} {892} (\bibinfo {year}
  {2020})},\ \bibinfo {note} {{DOI:10.1126/science.aba3993}}\BibitemShut
  {NoStop}%
\bibitem [{\citenamefont {Warken}\ \emph {et~al.}(2008)\citenamefont {Warken},
  \citenamefont {Rauschenbeutel},\ and\ \citenamefont
  {Bartholomaus}}]{Warken08}%
  \BibitemOpen
  \bibfield  {author} {\bibinfo {author} {\bibfnamefont {F.}~\bibnamefont
  {Warken}}, \bibinfo {author} {\bibfnamefont {A.}~\bibnamefont
  {Rauschenbeutel}},\ and\ \bibinfo {author} {\bibfnamefont {T.}~\bibnamefont
  {Bartholomaus}},\ }\bibfield  {title} {\bibinfo {title} {Fiber pulling
  profits from precise positioning - precise motion control improves
  manufacturing of fiber optical resonators},\ }\href@noop {} {\bibfield
  {journal} {\bibinfo  {journal} {Photonics Spectra}\ }\textbf {\bibinfo
  {volume} {3}},\ \bibinfo {pages} {73} (\bibinfo {year} {2008})}\BibitemShut
  {NoStop}%
\bibitem [{\citenamefont {Fujiwara}\ \emph {et~al.}(2011)\citenamefont
  {Fujiwara}, \citenamefont {Toubaru}, \citenamefont {Noda}, \citenamefont
  {Zhao},\ and\ \citenamefont {Takeuchi}}]{Fujiwara11}%
  \BibitemOpen
  \bibfield  {author} {\bibinfo {author} {\bibfnamefont {M.}~\bibnamefont
  {Fujiwara}}, \bibinfo {author} {\bibfnamefont {K.}~\bibnamefont {Toubaru}},
  \bibinfo {author} {\bibfnamefont {T.}~\bibnamefont {Noda}}, \bibinfo {author}
  {\bibfnamefont {H.-Q.}\ \bibnamefont {Zhao}},\ and\ \bibinfo {author}
  {\bibfnamefont {S.}~\bibnamefont {Takeuchi}},\ }\bibfield  {title} {\bibinfo
  {title} {Highly efficient coupling of photons from nanoemitters into
  single-mode optical fibers},\ }\href@noop {} {\bibfield  {journal} {\bibinfo
  {journal} {Nano Lett.}\ }\textbf {\bibinfo {volume} {1}},\ \bibinfo {pages}
  {4362–4365} (\bibinfo {year} {2011})},\ \bibinfo {note}
  {{DOI:10.1021/nl2024867}}\BibitemShut {NoStop}%
\bibitem [{\citenamefont {Yalla}\ \emph {et~al.}(2012)\citenamefont {Yalla},
  \citenamefont {Kien}, \citenamefont {Morinaga},\ and\ \citenamefont
  {Hakuta}}]{Yalla12}%
  \BibitemOpen
  \bibfield  {author} {\bibinfo {author} {\bibfnamefont {R.}~\bibnamefont
  {Yalla}}, \bibinfo {author} {\bibfnamefont {F.~L.}\ \bibnamefont {Kien}},
  \bibinfo {author} {\bibfnamefont {M.}~\bibnamefont {Morinaga}},\ and\
  \bibinfo {author} {\bibfnamefont {K.}~\bibnamefont {Hakuta}},\ }\bibfield
  {title} {\bibinfo {title} {Efficient channeling of fluorescence photons from
  single quantum dots into guided modes of optical nanofiber},\ }\href@noop {}
  {\bibfield  {journal} {\bibinfo  {journal} {Phys. Rev. Lett.}\ }\textbf
  {\bibinfo {volume} {109}},\ \bibinfo {pages} {063602} (\bibinfo {year}
  {2012})},\ \bibinfo {note} {{DOI:10.1103/PhysRevLett.109.063602}}\BibitemShut
  {NoStop}%
\bibitem [{\citenamefont {Skoff}\ \emph {et~al.}(2018)\citenamefont {Skoff},
  \citenamefont {Papencordt}, \citenamefont {Schauffert}, \citenamefont
  {Bayer},\ and\ \citenamefont {Rauschenbeutel}}]{Skoff18}%
  \BibitemOpen
  \bibfield  {author} {\bibinfo {author} {\bibfnamefont {S.~M.}\ \bibnamefont
  {Skoff}}, \bibinfo {author} {\bibfnamefont {D.}~\bibnamefont {Papencordt}},
  \bibinfo {author} {\bibfnamefont {H.}~\bibnamefont {Schauffert}}, \bibinfo
  {author} {\bibfnamefont {B.~C.}\ \bibnamefont {Bayer}},\ and\ \bibinfo
  {author} {\bibfnamefont {A.}~\bibnamefont {Rauschenbeutel}},\ }\bibfield
  {title} {\bibinfo {title} {Optical-nanofiber-based interface for single
  molecules},\ }\href@noop {} {\bibfield  {journal} {\bibinfo  {journal} {Phys.
  Rev. A}\ }\textbf {\bibinfo {volume} {197}},\ \bibinfo {pages} {043839}
  (\bibinfo {year} {2018})},\ \bibinfo {note}
  {{DOI:10.1103/PhysRevA.97.043839}}\BibitemShut {NoStop}%
\bibitem [{\citenamefont {Yonezu}\ \emph {et~al.}(2017)\citenamefont {Yonezu},
  \citenamefont {Wakui}, \citenamefont {Furusawa}, \citenamefont {Takeoka},
  \citenamefont {Semba},\ and\ \citenamefont {Aoki}}]{Yonezu17}%
  \BibitemOpen
  \bibfield  {author} {\bibinfo {author} {\bibfnamefont {Y.}~\bibnamefont
  {Yonezu}}, \bibinfo {author} {\bibfnamefont {K.}~\bibnamefont {Wakui}},
  \bibinfo {author} {\bibfnamefont {K.}~\bibnamefont {Furusawa}}, \bibinfo
  {author} {\bibfnamefont {M.}~\bibnamefont {Takeoka}}, \bibinfo {author}
  {\bibfnamefont {K.}~\bibnamefont {Semba}},\ and\ \bibinfo {author}
  {\bibfnamefont {T.}~\bibnamefont {Aoki}},\ }\bibfield  {title} {\bibinfo
  {title} {Efficient single-photon coupling from a nitrogen-vacancy center
  embedded in a diamond nanowire utilizing an optical nanofiber},\ }\href@noop
  {} {\bibfield  {journal} {\bibinfo  {journal} {Sci Rep}\ }\textbf {\bibinfo
  {volume} {7}},\ \bibinfo {pages} {12985} (\bibinfo {year} {2017})},\ \bibinfo
  {note} {{DOI:10.1038/s41598-017-13309-z}}\BibitemShut {NoStop}%
\bibitem [{\citenamefont {Ding}\ \emph {et~al.}(2020)\citenamefont {Ding},
  \citenamefont {Joos}, \citenamefont {Bach}, \citenamefont {Bienaimé},
  \citenamefont {Giacobino}, \citenamefont {Wu}, \citenamefont {A.Bramati},\
  and\ \citenamefont {Glorieux}}]{Ding20}%
  \BibitemOpen
  \bibfield  {author} {\bibinfo {author} {\bibfnamefont {C.}~\bibnamefont
  {Ding}}, \bibinfo {author} {\bibfnamefont {M.}~\bibnamefont {Joos}}, \bibinfo
  {author} {\bibfnamefont {C.}~\bibnamefont {Bach}}, \bibinfo {author}
  {\bibfnamefont {T.}~\bibnamefont {Bienaimé}}, \bibinfo {author}
  {\bibfnamefont {E.}~\bibnamefont {Giacobino}}, \bibinfo {author}
  {\bibfnamefont {E.}~\bibnamefont {Wu}}, \bibinfo {author} {\bibnamefont
  {A.Bramati}},\ and\ \bibinfo {author} {\bibfnamefont {Q.}~\bibnamefont
  {Glorieux}},\ }\bibfield  {title} {\bibinfo {title} {Nanofiber based
  displacement sensor},\ }\href@noop {} {\bibfield  {journal} {\bibinfo
  {journal} {Appl. Phys. B}\ }\textbf {\bibinfo {volume} {126}} (\bibinfo
  {year} {2020})},\ \bibinfo {note}
  {{DOI:10.1007/s00340-020-07452-1}}\BibitemShut {NoStop}%
\bibitem [{\citenamefont {Fogliano}\ \emph {et~al.}(2021)\citenamefont
  {Fogliano}, \citenamefont {Besga}, \citenamefont {Reigue}, \citenamefont
  {Heringlake}, \citenamefont {de~L\'{e}pinay}, \citenamefont {Vaneph},
  \citenamefont {Reichel}, \citenamefont {Pigeau},\ and\ \citenamefont
  {Arcizet}}]{Fogliano21}%
  \BibitemOpen
  \bibfield  {author} {\bibinfo {author} {\bibfnamefont {F.}~\bibnamefont
  {Fogliano}}, \bibinfo {author} {\bibfnamefont {B.}~\bibnamefont {Besga}},
  \bibinfo {author} {\bibfnamefont {A.}~\bibnamefont {Reigue}}, \bibinfo
  {author} {\bibfnamefont {P.}~\bibnamefont {Heringlake}}, \bibinfo {author}
  {\bibfnamefont {L.~M.}\ \bibnamefont {de~L\'{e}pinay}}, \bibinfo {author}
  {\bibfnamefont {C.}~\bibnamefont {Vaneph}}, \bibinfo {author} {\bibfnamefont
  {J.}~\bibnamefont {Reichel}}, \bibinfo {author} {\bibfnamefont
  {B.}~\bibnamefont {Pigeau}},\ and\ \bibinfo {author} {\bibfnamefont
  {O.}~\bibnamefont {Arcizet}},\ }\bibfield  {title} {\bibinfo {title} {Mapping
  the cavity optomechanical interaction with subwavelength-sized ultrasensitive
  nanomechanical force sensors},\ }\href@noop {} {\bibfield  {journal}
  {\bibinfo  {journal} {Phys. Rev. X}\ }\textbf {\bibinfo {volume} {11}},\
  \bibinfo {pages} {021009} (\bibinfo {year} {2021})},\ \bibinfo {note}
  {{DOI:10.1103/PhysRevX.11.021009}}\BibitemShut {NoStop}%
\bibitem [{\citenamefont {Drever}\ \emph {et~al.}(1983)\citenamefont {Drever},
  \citenamefont {Hall}, \citenamefont {Kowalski}, \citenamefont {Hough},
  \citenamefont {Ford}, \citenamefont {Munley},\ and\ \citenamefont
  {Ward}}]{Drever83}%
  \BibitemOpen
  \bibfield  {author} {\bibinfo {author} {\bibfnamefont {R.~W.~P.}\
  \bibnamefont {Drever}}, \bibinfo {author} {\bibfnamefont {J.~L.}\
  \bibnamefont {Hall}}, \bibinfo {author} {\bibfnamefont {F.~V.}\ \bibnamefont
  {Kowalski}}, \bibinfo {author} {\bibfnamefont {J.}~\bibnamefont {Hough}},
  \bibinfo {author} {\bibfnamefont {G.~M.}\ \bibnamefont {Ford}}, \bibinfo
  {author} {\bibfnamefont {A.~J.}\ \bibnamefont {Munley}},\ and\ \bibinfo
  {author} {\bibfnamefont {H.}~\bibnamefont {Ward}},\ }\bibfield  {title}
  {\bibinfo {title} {Laser phase and frequency stabilization using an optical
  resonator},\ }\href@noop {} {\bibfield  {journal} {\bibinfo  {journal} {Appl.
  Phys. B}\ }\textbf {\bibinfo {volume} {31}},\ \bibinfo {pages} {97} (\bibinfo
  {year} {1983})},\ \bibinfo {note} {{DOI:10.1007/BF00702605}}\BibitemShut
  {NoStop}%
\bibitem [{\citenamefont {Warken}(2007)}]{Warken07}%
  \BibitemOpen
  \bibfield  {author} {\bibinfo {author} {\bibfnamefont {F.}~\bibnamefont
  {Warken}},\ }\emph {\bibinfo {title} {Ultradünne Glasfasern als Werkzeug zur
  Kopplung von Licht und Materie}},\ \href@noop {} {\bibinfo {type} {{Ph.D.}
  thesis}},\ \bibinfo  {school} {University of Bonn, Germany} (\bibinfo {year}
  {2007}),\ \bibinfo {note}
  {https://hdl.handle.net/20.500.11811/3141}\BibitemShut {NoStop}%
\bibitem [{\citenamefont {van~de Hulst}(1981)}]{Hulst81}%
  \BibitemOpen
  \bibfield  {author} {\bibinfo {author} {\bibfnamefont {H.}~\bibnamefont
  {van~de Hulst}},\ }\bibinfo {title} {Light scattering by small particles}\
  (\bibinfo  {publisher} {Dover publication},\ \bibinfo {address} {New York},\
  \bibinfo {year} {1981})\ Chap.~\bibinfo {chapter} {15}, pp.\ \bibinfo {pages}
  {297--304},\ \bibinfo {edition} {1st}\ ed.\BibitemShut {Stop}%
\bibitem [{\citenamefont {Maurer}\ \emph {et~al.}(2021)\citenamefont {Maurer},
  \citenamefont {Gonzalez-Ballestero},\ and\ \citenamefont
  {Romero-Isart}}]{Maurer21}%
  \BibitemOpen
  \bibfield  {author} {\bibinfo {author} {\bibfnamefont {P.}~\bibnamefont
  {Maurer}}, \bibinfo {author} {\bibfnamefont {C.}~\bibnamefont
  {Gonzalez-Ballestero}},\ and\ \bibinfo {author} {\bibfnamefont
  {O.}~\bibnamefont {Romero-Isart}},\ }\bibfield  {title} {\bibinfo {title}
  {Quantum electrodynamics with a nonmoving dielectric sphere: Quantizing
  lorenz-mie scattering},\ }\href@noop {} {\bibfield  {journal} {\bibinfo
  {journal} {arXiv:2106.07975v2}\ } (\bibinfo {year} {2021})}\BibitemShut
  {NoStop}%
\end{thebibliography}%
%\end{thebibliography}

\end{document}